# First-principles calculations of the electronic, optical and elastic properties of $CdIn_2S_4$ spinel at the ambient and elevated pressure

V. Krasnenko, M.G. Brik[1]

*Institute of Physics, University of Tartu, Riia 142, Tartu 51014, Estonia*

**Abstract**

$CdIn_2S_4$ thiospinel was studied by means of the first principles calculations in both generalized gradient and local density approximations (GGA and LDA). The structural, electronic, optical and elastic properties were calculated in the pressure range from 0 to 10 GPa, below the pressure of phase transition for this compound. One of the main results of the paper is that the previously encountered in the literature controversy regarding the character of the $CdIn_2S_4$ band gap was resolved in favor of the indirect gap. Pressure coefficient of the band gap estimated as 0.071/0.063 eV/GPa (GGA/LDA) is in excellent agreement with the found in the literature experimental data of 0.076 or 0.069 eV/GPa. Calculated pressure dependence of the unit cell volume follows the experimental results very closely. Dependence of the interionic distances, lattice parameter and all elastic constants on pressure was calculated. Refined estimations of the Debye temperature for $CdIn_2S_4$ are given as 280 K (LDA) and 252 K (GGA). Elastic anisotropy of $CdIn_2S_4$ was visualized by plotting three-dimensional dependence of the Young modulus on a direction in the crystal lattice; it was established that the lowest Young moduli values are realized if the external stress is applied along the crystallographic axes.

## I. INTRODUCTION

Crystals with the spinel structure form a very large group of compounds with very diverse properties. They can be often met in nature occurring as pure or mixed minerals all over the globe, forming a very important part of the Earth's mantle and, as such, providing scientists with information about geological evolution of minerals and their behavior under extreme conditions, like extremely high pressure and/or temperature. Besides that, the spinels are also significant from the point of view of technological applications. Many representatives of this group of compounds are typical semiconductors with rather narrow band gap (this is true especially for spinels containing halogen atoms), whereas the oxygen-based spinels possess wider band gaps, which opens a possibility of easy and efficient doping with rare earth and transition metal ions. For example, $MgAl_2O_4$:$Co^{2+}$, $ZnGa_2O_4$:$Co^{2+}$, and $MgAl_2O_4$:$Ni^{2+}$ spinels were shown to be promising materials for the solid state lasers [1,2]. The spinel-based transparent ceramics were also designed recently for high-energy laser systems [3]. Highly efficient luminescent materials on the basis of the spinel compounds have been also reported [4, 5, 6]; application of spinels in diagnostics as fluorescence markers can be noted [7]. It should be also mentioned that the rare-earth ions doped spinels possess electrical conductivity, magnetic and semiconducting properties that can be potentially used in spintronics [8]. These examples (even being grouped into a very short list, which is not exhausting and far from being complete − but reviewing properties of all spinels is not the aim of the present paper) show undoubtedly great potential and importance of this group of materials.

---
[1] Corresponding author. E-mail: brik@fi.tartu.ee Phone +372 7374751

We shall focus our attention on one member of the thiospinel subgroup of a much larger spinels group (here *thio* refers to sulfur), namely on the cadmoindite (cadmium indium sulfide) with the chemical formula $CdIn_2S_4$. It is a chalcogenide semiconductor that belongs to the $AB_2X_4$ family of ternary compounds ($A$ and $B$ are metals in the II and III oxidation states, respectively (II - Cd, Mg, Mn, Zn; III - Al, Ga, In) or in the II and IV oxidation states (IV - Si, Ge, Sn), and $X$ usually stands for oxygen or a chalcogen [9, 10, 11, 12]).

The thiospinel subfamily forms an interesting group of materials for optoelectronic applications, which stem from their nonlinear optical properties. Their crystal structure is characterized by a partial degree of cation inversion, which leads to a high concentration of antisite defects. The concentration of these defects can be adjusted by varying external pressure, which makes thiospinels be materials of choice for defect engineering applications [13]. Furthermore, indium thiospinels have been recently proposed as ideal materials for photovoltaic cells operating with an intermediate band. The unusual physical properties of thiospinels are of interest because they present a new type of the metal−insulator transition [14, 15] and a pressure-induced superconductor−insulator transition [16].

In the last few years, there has been an increasing interest in understanding behavior of materials under compression based on calculations or measurements as it provides an insight into the nature of solid state theories and determines the values of fundamental parameters [17] . This applies especially to the compounds from the spinel family as well as to the thiospinel subfamily [9, 18, 19].

Cadmium indium sulfide is currently used for optoelectronic application as a photoconductor [19]. This compound is also considered as a promising candidate for intermediate-band (IB) formation, because its band gap lies in the region of optimum gaps for the implementation of an IB material [20]. Also, $CdIn_2S_4$ and a similar compound of $MgIn_2S_4$ attracts more and more attention because it is a visible-light-driven (VLD) photosensitive semiconductor that can be used in solar cells [21, 22, 23], hydrogen evolution [24] and organic dyes, such as MB (Methylene Blue) degradation [25], indicating that it may represent a new VLD photocatalyst for bacterial inactivation [ 26].

$CdIn_2S_4$ has been known for a long time being the subject of many experimental [22, 26, 27 , 28] and theoretical works [18, 19, 29 , 30, 31, 32]. Several reports on its optical properties, e.g. photoluminescence, absorption, reflectivity and Raman scattering, can be found in the literature [27, 33, 34, 35, 36, 37, 38]. The electronic properties have been also studied [33,39,40, 41, 42]. Furthermore, the elastic, electronic, and optical properties of the $MgIn_2S_4$ and $CdIn_2S_4$ spinels have been investigated by Semari et al [19]. More recently Bouhemadou et al. [18] have considered the influence of pressure on the thermodynamic properties of $MgIn_2S_4$ and $CdIn_2S_4$ compounds.

At the same time, although $CdIn_2S_4$ has been studied by several experimental and theoretical groups, still there exists some controversy regarding its physical properties, like band gap character (direct or indirect), pressure coefficients of the band gap value, Debye temperature etc.

Therefore, the main goal of this paper was set to clarify those questionable issues and gain more information on how some macroscopic parameters are affected by applied hydrostatic pressure, with subsequent comparison of the obtained results with corresponding experimental and theoretical data that are available in the literature. With this aim in view, we have carried out density functional calculations in order to model the pressure effects on the structural, elastic and thermodynamic properties of $CdIn_2S_4$.

The presented results are structured as follows: the details of the crystal structure and calculating settings are all given in the next section. Then the electronic and optical properties (including comparison with the experimental XPS spectra) are presented, followed by the pressure-affected changes of the physical properties of $CdIn_2S_4$. The paper is concluded with a summary of the most important results.

## II. CRYSTAL STRUCTURE AND DETAILS OF CALCULATIONS

As a typical member of the spinel group, the cadmium indium sulfide $CdIn_2S_4$ crystallizes in the cubic space group $Fd-\bar{3}mS$, No. 227. In this structure, the $A$ and $B$ atoms occupy the tetra- and octahedral sites, respectively [43]. Each $X$ anion is bonded to three $B$ cations located at the octahedral sites and with only one $A$ cation located at a tetrahedral sites [9]. In other words, each In cation is surrounded by six S anions and each Cd cation is surrounded by four S anions. Figure 1 shows one unit cell of $CdIn_2S_4$, with explicitly indicated coordination of cadmium, indium and sulfur ions.

All presented calculations were performed using the CASTEP module [44] of Materials Studio. The initial structural data needed for the optimization of the crystal structure were taken from Ref. [12]. To ensure consistency and better compatibility of the obtained results and drawn afterwards conclusions, two independent computational runs were made, using both the generalized gradient approximation (GGA) with the Perdew-Burke-Ernzerhof functional [45] and local density approximation (LDA) with the Ceperley-Alder-Perdew-Zunger (CA-PZ) functional [46, 47]. Since no defects have been considered, the calculations were performed for a primitive cell.

The plane-wave basis energy cutoff was chosen to be 350 eV. The Monkhorst-Pack scheme $k$-points grid sampling was set at 9×9×9 for the Brillouin zone. The convergence parameters were set as follows: total energy tolerance − $5 \times 10^{-6}$ eV/atom, maximum force tolerance 0.01 eV/nm, maximum stress 0.02 GPa, and maximum displacement $5 \times 10^{-4}$ Å. The electronic configurations were $4d^{10}5s^2$ for Cd, $4d^{10}5s^25p^1$ for In and $3s^23p^4$ for S. The effect of hydrostatic pressure on the structural parameters of $CdIn_2S_4$ at zero temperature has been investigated in the pressure range up to 10 Gpa with a step of 2 GPa. A complete optimization of the structural parameters were performed at each pressure.

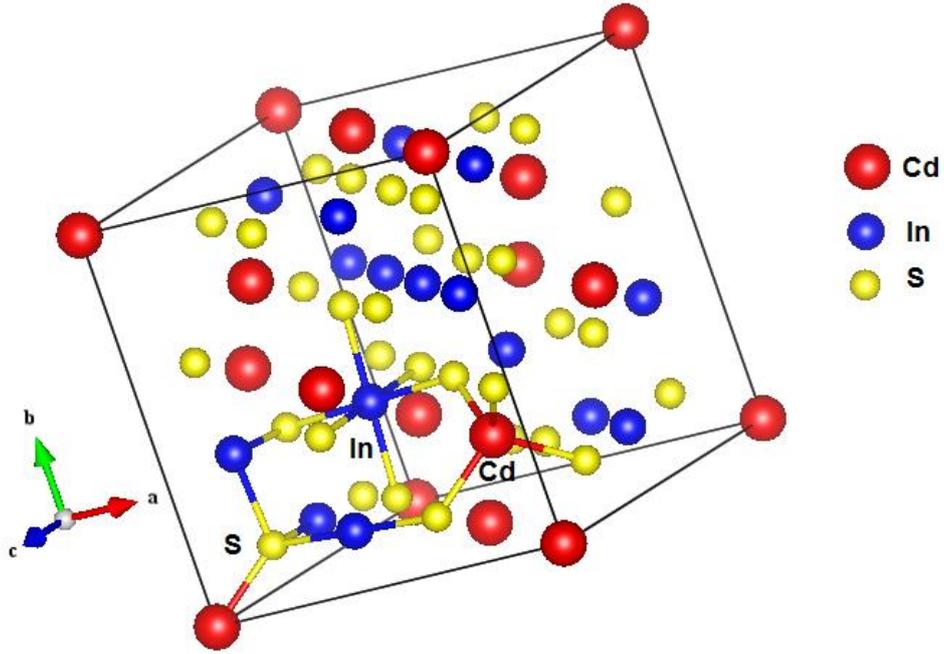

FIG. 1. (Color online) One unit cell of CdIn$_2$S$_4$. For simplicity, coordination is shown only for one Cd, one In, and one S atoms. Drawn with VESTA [48].

## III. RESULTS OF CALCULATIONS: STRUCTURAL, ELECTRONIC AND OPTICAL PROPERTIES AT THE AMBIENT PRESSURE

The structural parameters of CdIn$_2$S$_4$ along with the band gap values are summarized in Table I, which offers comparison of our calculated data with other corresponding results found in the literature. Agreement between the calculated in the present work and experimental lattice constant at ambient pressure is very good; the calculated results are also consistent with those reported in Refs. [9, 18, 19]. It can be noted that the GGA-obtained lattice constants is greater than the LDA-obtained one, which is a common observation for this type of calculations.

The calculated band gap is underestimated if compared to the experimental results, by about 1.2-1.4 eV. This is a usual underestimation, which can be handled later on (when calculating the optical properties) in a standard way by introducing the scissor operator producing a rigid up-ward shift of the conduction band.

Table I. Summary of structural and electronic properties of CdIn$_2$S$_4$

| | Calculated (this work) | | Experiment | Calculated (other works) |
|---|---|---|---|---|
| | LDA | GGA | | |
| Lattice constant, Å | 10.78673 | 11.06467 | 10.853[a]; 10.7979[b], 10.8378[c], | 10.7863[e], 11.107[f] |
| Unit cell volume, Å$^3$ | 1255.074 | 1354.613 | 1278.349[a], 1258.977[b], 1272.985[c] | 1254.924[e], 1370.220[f] |
| Cd-S distance, Å | 2.5044 | 2.5814 | 2.54[b] | |
| In-S distance, Å | 2.6028 | 2.6632 | 2.59[b] | |
| Band gap, eV | 1.212 | 1.047 | 2.35[c], 2.45[d], 2.02[g], 2.2[h] | 1.92[f], 1.12[f] |

[a] Ref. [34]

[b] Ref. [12]

[c] Ref. [13] (Direct band gap was reported)

[d] Ref. [49] (Direct band gap was reported)

[e] Ref. [18]

[f] Ref. [19]

[g] Ref. [26]

[h] Ref. [40]

Figure 2 presents the calculated band structure of CdIn2S$_4$, whereas the Brillouin zone for a primitive cell of CdIn$_2$S$_4$ indicating a path for the shown band structure is depicted in Figure 3.

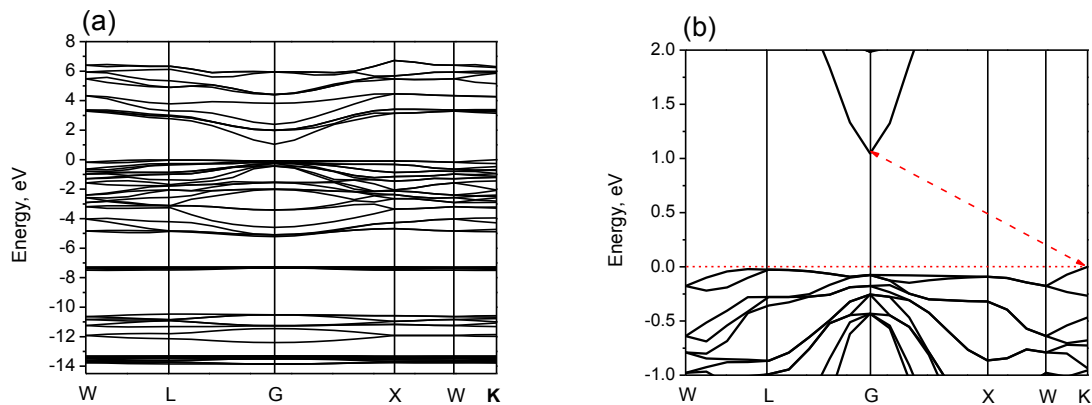

FIG. 2. (Color online) Calculated band structure of CdIn$_2$S$_4$ (a) and enlarged view of the band gap region (b).

The valence band is about 5 eV wide, whereas the conduction band width is about 6 eV. It should be pointed out that in the literature still exists a controversy regarding the character of the band gap. Some authors state that the CdIn$_2$S$_4$ is a direct band gap material [9, 13, 49], whereas some others suggest an indirect character of the band gap in this spinel [19, 21, 40]. As is evidenced by Figure 2b, our calculated band structure places CdIn$_2$S$_4$ into the group of indirect band gap compounds, because the maximum of the valence band and the minimum of the conduction band are realized at the different points of the Brillouin zone (K and G, respectively), in agreement with Ref. [19]. However, the difference between the energies of the top of the valence band at the K and G points is very small and equals to about 0.07 eV only, which may affect the results and final conclusions of the experimental studies of fundamental absorption of CdIn$_2$S$_4$.

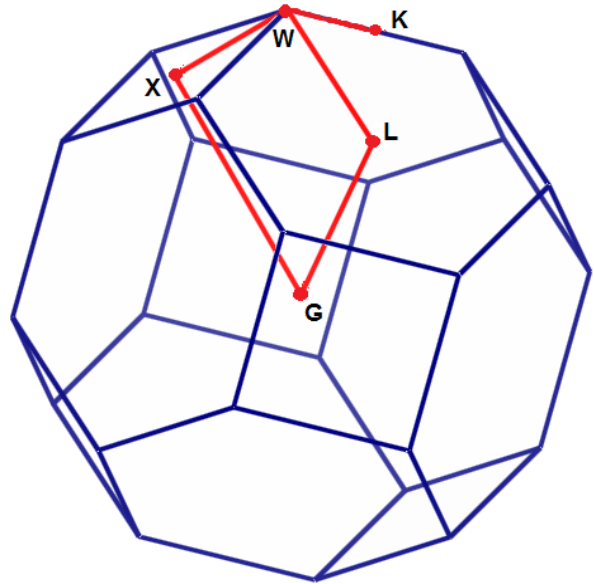

FIG. 3. (Color online) Brillouin zone for a primitive cell of CdIn$_2$S$_4$ and a path from the band structure diagram. The coordinates of the special points of the Brillouin zone are (in terms of the reciprocal lattice unit vectors: W(0.5, 0.25, 0.75), L(0.5, 0.5, 0.5), G(0, 0, 0), X(0.5, 0, 0.5), K(0.375, 0.375, 0.750).

The dispersion of the electronic states (Figure 2a) shows a low mobility of the holes in the conduction band and a higher mobility of the electrons in the conduction band, especially in the vicinity of the Brillouin zone center G.

The composition of the electronic bands can be analyzed using the element-resolved density of states (DOS) diagrams (Figure 4). The conduction band is mainly composed of the Cd and In 5s, 5p states, with an admixture of the S 3p states. It should be emphasized that the 5s states form the bottom of the conduction band, whereas the 5p states of In and the 3p states of S (they are due to the hybridization effects) tend to be located at the top of the conduction band. The S 3p states dominate in the upper part of the valence band; a minor contribution of the In and Cd 5s, 5p states to the valence band (especially its lower part, where the In 5s states are sharply peaked) can be also traced down. The completely filled Cd 4d states produce a narrow band at around -7.5 eV; the In 3d states are somewhat lower in energy at about -12 eV. Finally, the S 3s states are located somewhat deeper, stretching from -14 eV to -10 eV.

A reliable check of the calculated electronic band structure and DOS diagrams comes from the experimental X-ray photoelectron spectroscopy (XPS). The experimental XPS spectra of CdIn$_2$S$_4$ were reported in Refs. [50, 51]. They are compared with the sum of the partial density of states (PDOS) in Figure 5.

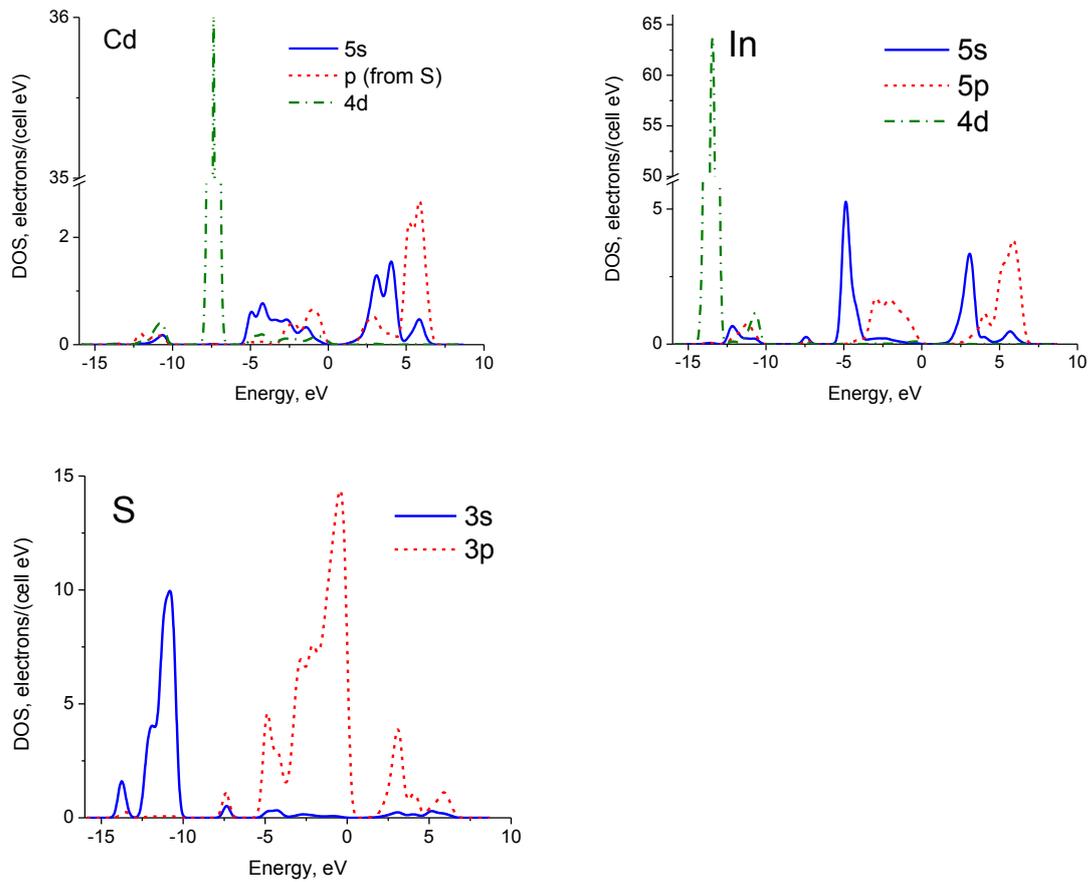

FIG. 4. (Color online) Calculated density of states (DOS) diagrams for CdIn$_2$S$_4$.

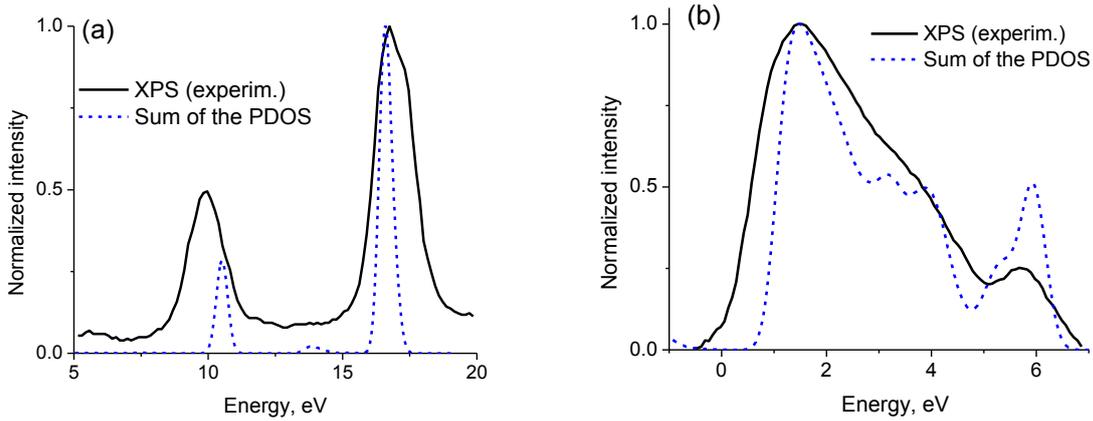

FIG. 5. (Color online) Comparison of the experimental XPS spectra [50] with the results of the DOS calculations for $CdIn_2S_4$.

The energy scales of the XPS spectra are shifted, as it is done usually, to reach the best agreement with the relative features of the DOS distributions, which, in their turn, are drawn in respect to the valence band top. The sum of the PDOS was calculated by adding the contributions of individual chemical elements (Cd, In, S), each of which was multiplied by the number of atoms of a particular element in a chemical formula (1 for Cd, 2 for In, 4 for S). The spectra were normalized by the intensity of the most prominent feature in the considered spectral region.

The highest in energy peak in the experimental XPS spectrum is located at about 17 eV; it is well reproduced by the In 4d states distribution; the Cd 4d states, which are located higher in energy, are in charge of the 10 eV peak in the XPS spectrum (Figure 5a). The width of the low-energy part of the XPS spectrum (Figure 5b) is very close to the width of the calculated valence band. The contribution of the sulfur 3p states determines the overall structure of this range of the experimental spectrum: the most intensive contribution (at about 2 eV) comes from the S 3p states located closer to the top of the valence band; other well-seen experimental features at about 2.5 and 5 eV follow the peculiar behavior of the 3p states distribution closer to the bottom of the valence band.

We also report here the calculated values of the refractive index in the limit of infinite wavelengths are 2.25 (GGA) and 2.33 (LDA), which is in good agreement with the experimental result of 2.55 [33] and other calculated values of 2.4 [19] and 2.625, 2.893 [29].

Figure 6 exhibits a cross-section of the electron density difference in the plane, whose normal is along the $(-\frac{\sqrt{2}}{2}, 0, \frac{\sqrt{2}}{2})$ direction in the crystal lattice. As follows from Fig. 6, the Cd-S bond is more covalent than the In-S one, because the electron density distribution around In ions is more spherical and does not exhibit any directional dependence like that one for Cd. Such a conclusion is also confirmed by the PDOS diagrams (Fig. 4), which show the presence of the p electron density (transferred from the sulfur ions) on the cadmium ions.

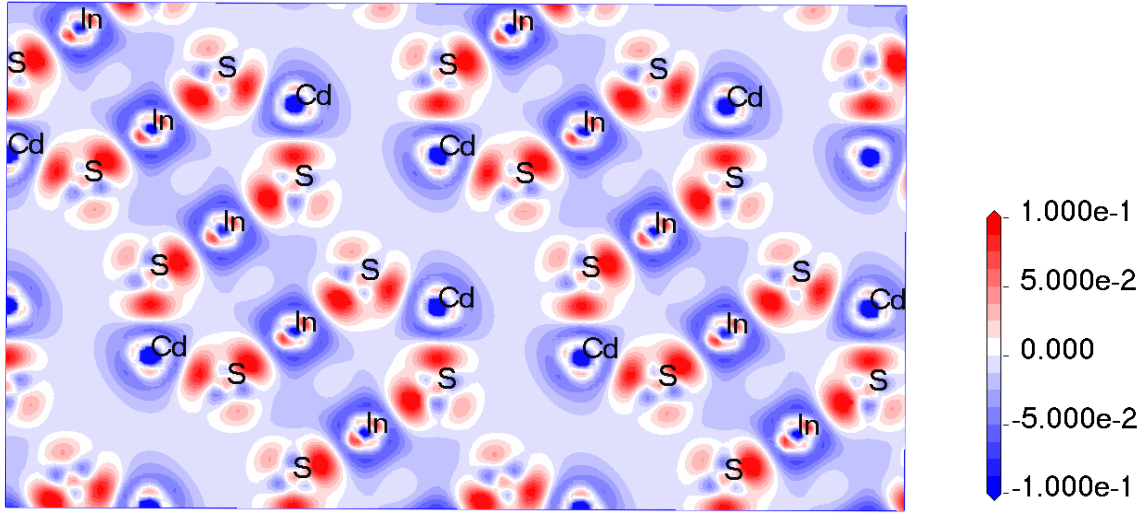

FIG. 6. (Color online) Cross section of the electron density difference for $CdIn_2S_4$. |The scale is in electrons/$\mathring{A}^3$.

## IV. RESULTS OF CALCULATIONS: ELASTIC AND THERMODYNAMIC PROPERTIES

The optimized crystal structure of the studied cadmoindite was used as an initial structure to calculate components of the elastic constants tensor; all calculating settings were kept the same as for the geometry optimization. Since $CdIn_2S_4$ is a cubic crystal, only three non-zero elastic constants $C_{11}$, $C_{12}$, and $C_{44}$ are needed to describe completely the response of such a material to external stresses. The bulk modulus $B$ for a cubic crystal can be calculated as $B =(C_{11} + 2C_{12})/3$. The calculated at the ambient pressure values of $C_{ij}$ and $B$ are all collected in Table II. As a common observation, it can be noted that the LDA-calculated elastic constants are greater than their GGA counterparts. The stability criteria for a cubic crystal $\left(\frac{1}{3}(C_{11} + 2C_{12}) > 0, C_{44} > 0, \frac{1}{2}(C_{11} - C_{12}) > 0\right)$ [52] are all fulfilled, thus meaning that the cubic phase of $CdIn_2S_4$ is stable.

Very few experimental data on the $CdIn_2S_4$ elastic constants and bulk modulus exist in the literature. In particular, the $C_{ij}$ values reported here can be compared only to other theoretical estimations and only one experimental study (Table II). The calculated values of bulk modulus at zero pressure (especially calculated by LDA) are in good agreement with the experimental result from Ref. [13].

Table II. Summary of elastic properties of $CdIn_2S_4$

|          | Calc. (this work) | | Calc. (other works) | Exp. |
|----------|------|--------|---------------------------|--------------------|
|          | GGA  | LDA    |                           |                    |
| $C_{11}$ | 77.55 | 100.01 | 96[a], 102.46[b]          | 121.5[c]           |
| $C_{12}$ | 55.51 | 65.73  | 69[a], 49.13[b]           | 24.6[c]            |
| $C_{13}$ | 34.59 | 38.21  | 38[a], 21.46[b]           | 25.7[c]            |
| $B$      | 62.86 | 77.15  | 78[a], 66.90[b]           | 78 [d]             |

[a] Ref. [18]
[b] Ref. [19]
[c] Ref. [53]
[d] Ref. [13]

Agreement between our and other calculated values of $C_{ij}$ is good. We also note that the experimental value of $C_{12}$ reported in Ref. [53] seems to be considerably underestimated.

In spite of the cubic crystal structure of $CdIn_2S_4$, its elastic properties exhibit certain anisotropy, which can be analyzed by plotting the three-dimensional dependence of the Young modulus on the direction in the crystal lattice described by the following equation [54]:

$$E(\vec{n}) = \frac{1}{S_{11} - \beta_1(n_1^2 n_2^2 + n_1^2 n_3^2 + n_2^2 n_3^2)}, \text{ where } \beta_1 = 2S_{11} - 2S_{12} - S_{14} \qquad (1)$$

The $S_{ij}$ entries in Eq. (1) are the elastic compliance constants, which form the matrix inverse to the matrix of the elastic constants $C_{ij}$, and $n_1$, $n_2$, $n_3$ are the direction cosines specifying the direction in the crystal lattice. In the case of a perfect elastically isotropic medium Eq. (1) would describe a sphere. But in the case of $CdIn_2S_4$ such a surface deviates from the spherical shape, as is evidenced by Figure 7. It turns out to have the shape of a cube with depressions at the center of each face, which indicates that the smallest Young moduli values are realized if the stress is applied along the crystallographic axes, whereas the (111) direction in the crystal lattice is characterized by the greatest Young moduli values.

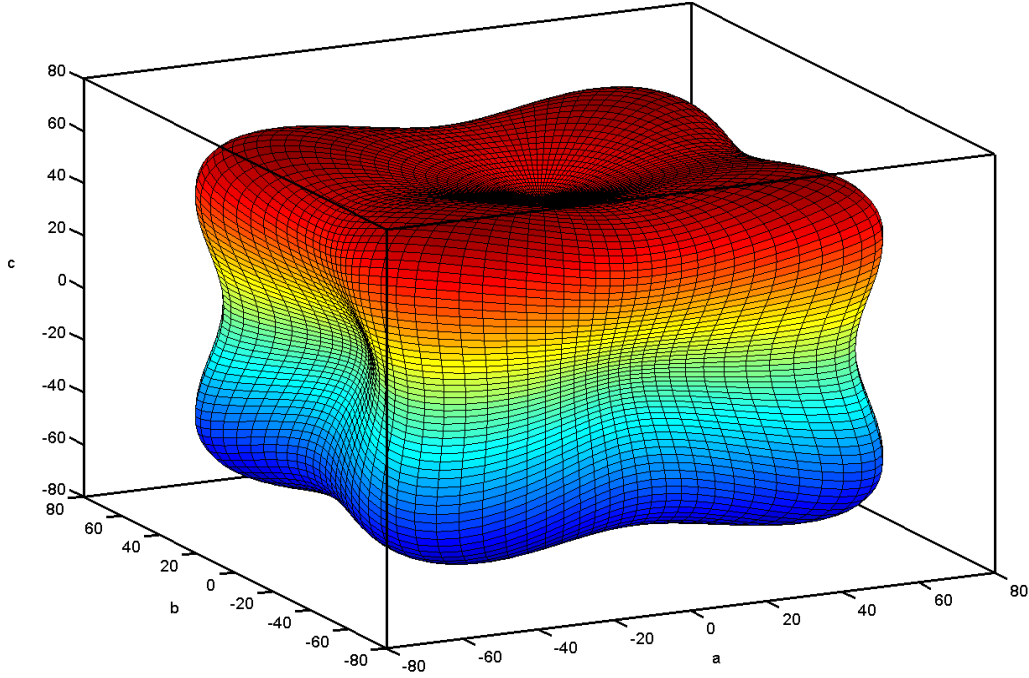

FIG. 7. (Color online) Visualization of the Young modulus surface (Eq. (1)) for $CdIn_2S_4$. The axes units are GPa.

After the elastic constants are calculated, one can proceed with estimations of the sound velocities in a solid and its Debye temperature $\Theta_D$, related to the maximum phonon energy in the Debye model, using the following equation [55]:

$$\Theta_D = \frac{h}{k}\left[\frac{3n}{4\pi}\frac{N_A\rho}{M}\right]^{1/3} v_m,$$  (2)

where $h$ and $k$ are the Planck's and Boltzmann's constants, respectively, $N_A$ is the Avogadro's number, $\rho$ is the crystal's density, $M$ is the molecular weight, $n$ denotes the number of atoms per one formula unit. The $v_m$ entry is the mean sound velocity expressed in terms of the longitudinal $v_l$ and transverse $v_t$ sound velocities as [56]

$$v_m = \left[\frac{1}{3}\left(\frac{2}{v_t^3} + \frac{1}{v_l^3}\right)\right]^{-1/3},$$  (3)

which, in their turn, are calculated as [57] $v_l = \sqrt{\frac{3B+4G}{3\rho}}$, $v_t = \sqrt{\frac{G}{\rho}}$, with $B$ being the bulk modulus and $G = (G_V + G_R)/2$ the isotropic shear modulus estimated as the average value of the Voigt's shear modulus $G_V$ (an upper limit for $G$ values) and the Reuss's shear modulus $G_R$ (a lower limit for $G$ values). The $G_V$ and $G_R$ values are expressed in terms of the elastic constants $C_{ij}$:

$$G_V = \frac{C_{11}-C_{12}+3C_{44}}{5}, \frac{5}{G_R} = \frac{4}{C_{11}-C_{12}} + \frac{3}{C_{44}} \qquad (4)$$

Using Eqs. (2) – (4), we estimated the Debye temperature as 280 K (LDA) and 252 K (GGA). There are no experimental estimations of the Debye temperature for CdIn$_2$S$_4$; the calculated values of 320 K [18] and 313 K [19] were reported; they are reasonably close to our results with the deviation presumably caused by different computational settings employed.

Table III summarizes the shear moduli, sound velocities and Debye temperature for CdIn$_2$S$_4$ calculated at ambient pressure. The ambient pressure values are compared to the calculated results from Refs. [18, 19].

As a fundamental parameter, the Debye temperature correlates with many physical properties of solids, such as specific heat, elastic constants, and melting temperature. At low temperatures the vibrational excitations arise solely from the acoustic vibrations. Hence, at low temperatures the Debye temperature calculated from the elastic constants is the same as that one determined from the specific heat measurements [58]. It should be noted that the Debye temperature obtained from the elastic constants is always somewhat higher than that one obtained from electrical resistivity studies. The discrepancy may be partly ascribed to its temperature dependence [58].

Table III. Calculated values of the shear moduli, density, sound velocities and Debye temperature for CdIn$_2$S$_4$ at ambient pressure. The calculated results from Refs. [18, 19] are given in italic.

| | $G_V$ (GPa) | $G_R$ (GPa) | $G$ (GPa) | Density (kg/m$^3$) | $v_l$ (m/s) | $v_t$ (m/s) | $v_m$ (m/s) | $\Theta_D$ (K) |
|---|---|---|---|---|---|---|---|---|
| LDA | 29.78 | 25.61 | 27.70 | 4977.78 | 4787 | 2359 | 2648 | 280 |
| GGA | 25.16 | 18.64 | 21.90 | 4612.00 | 4468 | 2179 | 2448 | 252 |
| | | | *25*[a] | *4978*[a] | *4730*[a] | *2240*[a] | *3030*[a] | *320*[a] |
| | | | *23.41*[b] | *4560*[b] | *4640*[b] | *2270*[b] | *3050*[b] | *313*[b] |

[a]Ref. [18]
[b]Ref. [19]

We note that the values of $v_l$ and $v_t$ from the present work and those from Refs. [18, 19] are quite close, although the $v_m$ values differ considerably, which, in turn, causes the discrepancy in the calculated Debye temperature values. Trying to understand this difference, we noticed that most probably it was due to improper application of Eq. (3) in Refs. [18, 19]: interchange of $v_l$ and $v_t$ in Eq. (3) would yield exactly those data obtained by the authors of those references.

Validity of our estimation of the Debye temperature is indirectly supported by the experimental values of the Debye temperature of 153 K for $MnIn_2S_4$ [49], 205 K for $CuV_2S_4$ [59] and 230 K for $CuRh_2S_4$ [60], which are much closer to our data than to the values of 313 or 320 K from Refs. [18, 19].

# V. RESULTS OF CALCULATIONS: PRESSURE EFFECTS ON THE STRUCTURAL, OPTICAL, AND ELASTIC PROPERTIES

For an analysis of influence of pressure on the $CdIn_2S_4$ properties its crystal structure has been optimized at various hydrostatic pressures in the range from 0 to 10 GPa with a step of 2 GPa. The upper limit of the pressure variation was restricted by 10 GPa, since there is a pressure-induced phase transition at about 10 GPa, as reported in Ref. [9].

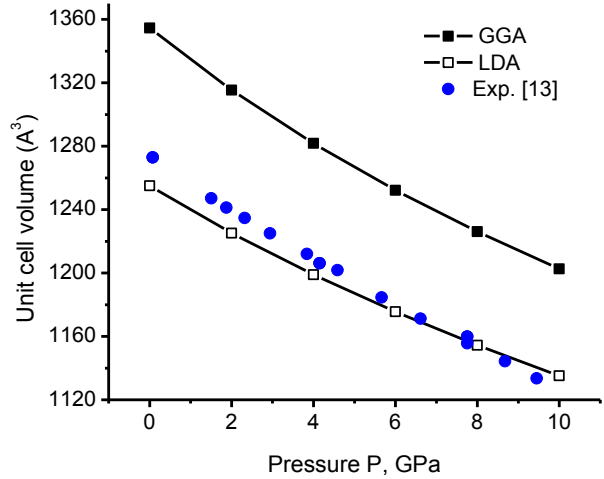

FIG. 8. (Color online) Pressure dependence of the unit cell volume of $CdIn_2S_4$.

Figure 8 shows how the unit cell volume of $CdIn_2S_4$ varies with pressure in comparison with the experimental data from Ref. [13]. As seen from the Figure, the LDA-calculated results (empty squares) follow very closely the experimental data, whereas the GGA-data tend to overestimate the experimental unit cell volume by about 5-6 %.

Another way to use the data presented in Figure 8 is to fit the corresponding data points to the Murnaghan equation of state [61]

$$\frac{V}{V_0} = \left(1 + P\frac{B\prime}{B}\right)^{-\frac{1}{B\prime}} \quad (5)$$

here $V_0$ is the volume at the ambient pressure, $B$ and $B' = \mathrm{d}B/\mathrm{d}P$ are the bulk modulus and its pressure derivative, respectively. Application of Eq. (5) to the data from Figure 8 yields the following values: $B$=63.8 GPa, $B'$=4.45 (GGA, this work), $B$=78.7 GPa, $B'$=4.51 (LDA, this work), $B$=74.6 GPa, $B'$=1.67 (experiment, [13]). These values are highly consistent with those from Table II.

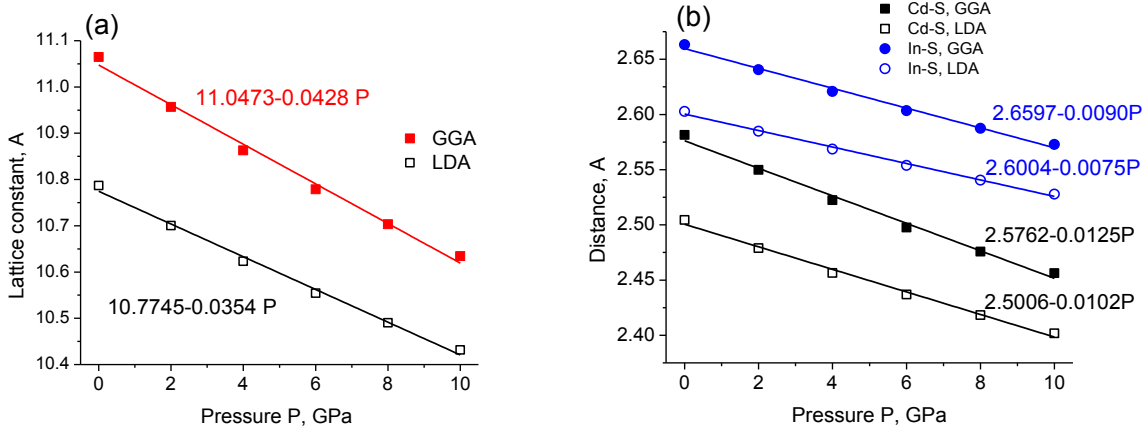

FIG. 9. (Color online) Pressure dependence of the lattice constant (a) and Cd-S, In-S distances (b) for CdIn$_2$S$_4$.

Useful information about the response of individual chemical bonds to the applied pressure can be gained by plotting the variation of the lattice constant and interatomic distances versus pressure (Figure 9). The lattice constant decreases with the pressure coefficients of -0.0428 Å/GPa (GGA) and -0.0354 Å/GPa (LDA). The Cd-S chemical bond shrinks with the pressure coefficient of -0.0125 Å/GPa (GGA) and -0.0102 Å/GPa (LDA). The In-S chemical bond decreases somewhat slower, with the pressure coefficient of -0.0090 Å/GPa (GGA) and -0.0075 Å/GPa (LDA).

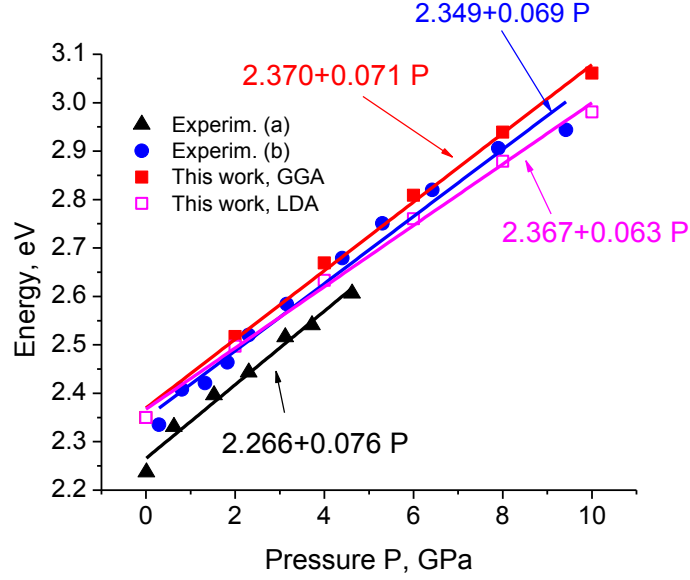

FIG. 10. (Color online) Pressure dependence of the calculated and experimental band gap for CdIn$_2$S$_4$. Experimental data were taken from Refs. [9, 62].

Decrease of the interatomic distances induced by pressure can be manifested in the blue shift of the absorption edge. The calculated band gaps, with the corresponding scissor operator 1.303/1.138 eV (GGA/LDA) taken into account, in comparison with the experimental data [9, 62] are presented in Figure 10. The band gap value increases linearly with pressure; its pressure coefficient is 0.076 or 0.069 eV/GPa (as extracted from the experimental data), whereas the GGA/LDA calculated pressure coefficients are 0.071/0.063 eV/GPa, in very good agreement with the experimental results.

Increase of the calculated band gap, as shown in Figure 10, also leads to the blue shift of the absorption spectrum (Figure 11). What is also worthwhile noting is that intensity of the calculated absorption bands, which correspond to the band-to-band transitions, increases with pressure, indicating enhanced hybridization effects between the electronic states of Cd and In, on one hand, and S, on the other hand.

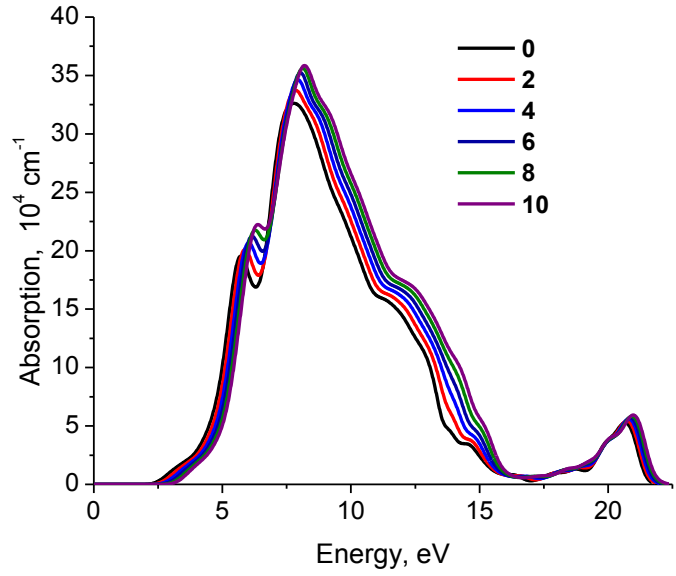

FIG. 11. (Color online) Pressure dependence of the calculated absorption spectrum for CdIn$_2$S$_4$. The values of pressure in GPa are given in the figure.

Elastic constants also increase with pressure (Figure 12), except for the $C_{44}$ constant, which is practically not changing at all. The linear approximations of all elastic constants as functions of pressure are presented in Table IV. It can be noted that the coefficient of linear approximation for the bulk modulus $B$ from Table 4 is very close to the $B'$ value obtained after application of the Murnaghan equation of state.

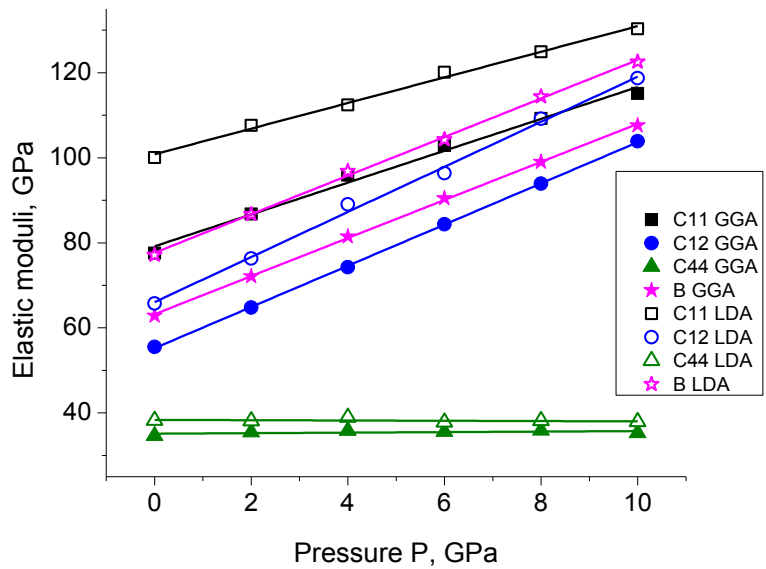

FIG. 12. (Color online) Pressure dependence of the elastic constants for CdIn$_2$S$_4$.

Table IV. Linear dependencies of elastic constants on pressure $P$ for $CdIn_2S_4$ (Fig. 10). Pressure $P$ and the calculated results are all expressed in GPa.

|          | GGA              | LDA               |
|----------|------------------|-------------------|
| $C_{11}$ | 79.18 + 3.75 $P$ | 100.81 + 3.01 $P$ |
| $C_{12}$ | 55.21 + 4.85 $P$ | 66.05 + 5.30 $P$  |
| $C_{44}$ | 35.12 + 0.06 $P$ | 38.36 - 0.03 $P$  |
| $B$      | 63.20 + 4.48 $P$ | 77.64 + 4.54 $P$  |

Another interesting result is that the effective Mulliken charges of all ions differ considerably from the formal charges expected from the chemical formula. They are (in the proton charge units, GGA/LDA values are given): 0.77/0.71 for Cd; 0.74/0.65 for In; -0.56/-0.50 for S. All charges are slightly increased with pressure, to reach the following values: 0.78/0.72 for Cd; 0.75/0.66 for In; -0.57/-0.51 for S, which indicates of the pressure-induced charge transfer from metal ions to sulfur.

## VI. CONCLUSIONS

In the present paper we reported the results of the first principles calculations of the structural, electronic, optical and elastic properties of $CdIn_2S_4$ thiospinel. After successfully performed optimization of the crystal lattice performed by using two approximations – GGA and LDA, the calculations of the electronic properties have shown this compound to possess the indirect band gap. We also report a refined value of the Debye temperature as 280 K (LDA) and 252 K (GGA). The pressure effects on the structure and electronic properties of $CdIn_2S_4$ were studied in the pressure range until 10 GPa. The calculated unit cell volumes using the LDA approach excellently follow the experimental data. The pressure coefficient of the band gap, calculated here for the first time, excellently follows the experimental data. The values of the elastic constants, bulk modulus and its pressure derivative were all evaluated. Elastic anisotropy of $CdIn_2S_4$ was assessed by visualizing the three-dimensional surface of the directional dependence of the Young modulus. It was demonstrated that the lowest Young moduli values (the highest compressibility) of $CdIn_2S_4$ is realized along the crystallographic axes, whereas the (111) direction is characteristic for the greatest Young's moduli values.

## ACKNOWLEDGEMENTS

This study was supported by the European Union through the European Regional Development Fund (Centre of Excellence "Mesosystems: Theory and Applications", TK114), European Internationalisation Programme DoRa and LUMINET - European Network on Luminescent Materials. Fruitful discussions with Prof. A. Suchocki (Institute of Physics, Polish Academy of Sciences), bilateral project between the Estonian and Polish Academies of Sciences in the years 2013-2015 and the project number DEC-2012/07/B/ST5/02080 of National Science Center of Poland are all gratefully acknowledged. We also thank Dr. G.A. Kumar (University of Texas at San Antonio) for allowing to use the Materials Studio package and Dr. Oleg Khyzhun (Institute for Problems of Materials Science, National Academy of Sciences of Ukraine) for help with some archival literature.